\documentclass[structabstract]{aa}  

\usepackage{graphicx}
\usepackage{txfonts}
\usepackage{natbib}
\bibpunct{(}{)}{;}{a}{}{,}
\newcommand{\kms}{\,{\rm km\,s^{-1}}}
\newcommand{\hMpc}{\, h^{-1}\, \mathrm{Mpc}}

\newcommand{\hMs}{\, h^{-1}\, M_{\odot}}
\def\spose#1{\hbox to 0pt{#1\hss}}
\def\lta{\mathrel{\spose{\lower 3pt\hbox{$\mathchar"218$}} \raise 2.0pt\hbox{$\mathchar"13C$}}}
\def\gta{\mathrel{\spose{\lower 3pt\hbox{$\mathchar"218$}} \raise 2.0pt\hbox{$\mathchar"13E$}}}

\begin{document}
\title{Wavelet analysis of baryon acoustic structures in the galaxy distribution}
\titlerunning{Wavelet analysis of baryon acoustic structures}

\author{
Pablo~Arnalte-Mur \inst{\ref{inst1},\ref{inst2},\ref{inst8}} \and Antoine~Labatie \inst{\ref{inst3}} \and Nicolas~Clerc \inst{\ref{inst3}} \and Vicent~J.~Mart\'{\i}nez \inst{\ref{inst1},\ref{inst4}} \and Jean-Luc~Starck \inst{\ref{inst3}} \and Marc~Lachi\`eze-Rey \inst{\ref{inst5}} \and Enn~Saar \inst{\ref{inst6}} \and Silvestre~Paredes \inst{\ref{inst7}}
}
\authorrunning{Arnalte-Mur et al.}

\institute{
Observatori Astron\`omic, Universitat de Val\`encia, Apartat de Correus 22085, E-46071 Val\`encia, Spain \label{inst1}
\and
Institut de F\'isica Corpuscular (CSIC-UVEG), Val\`encia, Spain \label{inst2}
\and
Institute for Computational Cosmology, Department of Physics, Durham University, South Road, Durham DH1 3LE, UK 
\email{pablo.arnalte-mur@durham.ac.uk}\label{inst8}
\and
Laboratoire AIM (UMR 7158), CEA/DSM-CNRS-Universit\'e Paris Diderot, IRFU, SEDI-SAP, Service d'Astrophysique, Centre de Saclay,  F-91191 Gif-Sur-Yvette Cedex, France \label{inst3}
\and
Departament d'Astronomia i Astrof\'{\i}sica, Universitat de Val\`encia, 46100-Burjassot, Val\`encia, Spain \label{inst4}
\and
Astroparticule et Cosmologie (APC), CNRS-UMR 7164,  Universit\'e Paris 7, Denis Diderot 10, rue Alice Domon et L\'eonie Duquet,  F-75205 Paris Cedex 13, France \label{inst5}
\and
Tartu Observatoorium, EE-61602 T\~oravere, Estonia \label{inst6}
\and
Departamento de Matem\'atica Aplicada y Estad\'{\i}stica, Universidad Polit\'ecnica de Cartagena, C/Dr. Fleming s/n, 30203 Cartagena, Spain \label{inst7}
}

\date{Received XXX; accepted YYY}


\abstract
{Baryon Acoustic Oscillations (BAO) are a feature imprinted in the density field by acoustic waves travelling in the plasma of the early universe. 
Their fixed scale can be used as a standard ruler to study the geometry of the universe.}
{BAO have been previously detected using correlation functions and power spectra of the galaxy distribution. In this work, we present a new method for the detection of the real-space structures associated with this feature. These baryon acoustic structures are spherical shells with a relatively small density contrast, surrounding high density central regions.}
{We design a specific wavelet adapted to the search for shells, and exploit the physics of the process by making use of two different mass tracers, introducing a specific statistic  to detect the BAO features. We show the effect of the BAO signal in this new statistic when applied to the $\Lambda$ -- Cold Dark Matter ($\Lambda$CDM) model, using  an analytical approximation to the transfer function.We confirm the reliability and stability of our method by using cosmological $N$-body simulations from the MareNostrum Institut de Ci\`encies de l'Espai (MICE).}
{We apply our method to the detection of BAO in a galaxy sample drawn from the Sloan Digital Sky Survey (SDSS). We use the `Main' catalogue to trace the shells, and the Luminous Red Galaxies (LRG) as tracers of the high density central regions. Using this new method, we detect, with a high significance, that the LRGs in our sample are preferentially located close to the centres of shell-like structures in the density field, with characteristics similar to those expected from BAOs. We show that stacking selected shells, we can find their characteristic density profile.}
{We have delineated a new feature of the cosmic web, the BAO shells. As these are real spatial structures, the BAO phenomenon can be studied in detail by examining those shells.}


   \maketitle
%

\section{Introduction}
\label{sec:intro} 

Before recombination, the energy of photons is high enough to avoid the formation
of neutral hydrogen atoms. This means that baryons and photons are
coupled through Compton scattering and electromagnetic interaction between
protons and electrons, forming a plasma. In this 
fluid two phenomena act in opposite
directions: gravitational forces tend to compress the plasma around high density
regions, while radiation pressure tends to dilute any such over-density.
The combination of both in the presence of any initial inhomogeneity give rise to
acoustic waves propagating in the baryon-photon plasma. This phenomenon ends
abruptly at the epoch of recombination, when the temperature drops enough to allow
hydrogen atoms to form, and therefore radiation decouples from the baryons.

Baryon acoustic oscillations (BAO) are therefore due to the propagation of these 
sound waves in the baryon-photon plasma in the early universe \citep{pee70a,hu97a,eis98a,bas09a}. 
Any primordial over-density in the early universe produces a spherical acoustic wave 
in the baryon-photon plasma, travelling outwards: the radiation pressure drags the baryons that are coupled to the photons, and compensates the gravity force that pulls all matter towards the centre. Dark matter, however, is totally decoupled from the photons, and therefore its density at the centre continues growing. About $380,000$ years after the Big Bang, temperature drops so that photons and baryons decouple, and the scale of the baryon shells freezes. After this time, both the central over-density and the shell grow gravitationally, accreting both dark matter and baryons. The result at late times is a large over-density at the position of the original perturbation, surrounded by a faint spherical shell at a fixed co-moving scale \citep{eis07a}. 

The BAO scale is fixed by the sound horizon at decoupling: it is the distance that the expanding acoustic shells can travel before decoupling. It has been accurately measured by the study of the anisotropies in the Cosmic Microwave Background (CMB)  to be \citep{kom08a} $r_s = 153.3 \pm 2.0 \, \mathrm{Mpc} = 110.4 \pm 1.4 \hMpc$ (where we take $h = 0.72$, \citealp{fre01a})\footnote{$h$ is the Hubble constant in units of $100 \kms\, \mathrm{Mpc}^{-1}$}.  Therefore, this scale, once measured, could be used as a standard ruler to measure the Hubble expansion rate with redshift $H(z)$ and the angular diameter distance $D_A(z)$ \citep{coo01a,bla03c,seo05a}.

The BAO should appear as a series of damping wiggles in the matter power spectrum, with the locations of the peaks and throughs in $k$-space being a function of  $r_s$ and other cosmological parameters \citep{eis98a}.  All the harmonics sum up to the same peak in the galaxy correlation function $\xi(r)$ at the scale $r_s$, and therefore it could seem more appropriate to use this statistic for the detection of the BAO feature on the available galaxy redshift surveys encompassing large volumes of the universe \citep{san08b}.

The first  detection (claiming a $\sim 3\sigma$ level)  was reported in the analysis of the correlation function \citep{eis05a} 
of the  Sloan Digital Sky Survey (SDSS) \citep{yor00a} Luminous Red Galaxies (LRG) sample \citep{eis01a}, and later in the power spectrum \citep{col05a} of the 2-degree Field Galaxy Redshift Survey (2dFGRS) \citep{col01a}. But certainly this is a controversial topic. \citet{cab10a} are not finding such level of detection using a data set twice as large in volume and in number of galaxies. They do not claim this result to be in contradiction with the standard $\Lambda$CDM model, but to be a consequence of insufficient data. One of the arguments in \citet{cab10a} is the fact that mixing model selection with parameter determination can lead to some confusion in the interpretation of the  results and their significance. Different authors are using different criteria to assess the significance of ther BAO detection. For example, when \citep{eis05a} affirm that the baryon signature was detected at 3.4 $\sigma$ (or at 3.0 $\sigma$ when including only data points between 60 and 180 $h^{-1}$ Mpc) they are comparing their results of the SDSS-LRG correlation function with the expected for the best-fit pure CDM model and  different BAO models. The best BAO detection up to now \citep{per10a} was obtained studying the combined power spectrum of LRG and `Main' \cite{str02a} samples of SDSS, together with the 2dFGRS sample, and is at the $\sim 3.6\sigma$ level.  The authors explicitly  state that since this number is obtained comparing to an arbitrary smooth model,  the significance cannot be directly compared with the one reported  in \citep{eis05a}.  This is a clear example of different authors using different ways to assess the significance of their results that in practice are not comparable. \citet{hut06a} calculated the redshift space power spectrum of the  SDSS-LRG sample drawn from the Data Release 4. He concludes that BAO models are favored by $3.3 \sigma$ over the corresponding models without any oscillatory behavior in the power spectrum. 

\citet{per07a} detected BAOs in the clustering of the combined 2dFGRS and SDSS main galaxy 
samples, and use their measurements to constrain cosmological models, in particular a given combination of
the angular diameter distance $D_A(z)$ and the Hubble parameter $H(z)$. \citet{cab08a,cab08b} studied the LRGs anisotropic redshift space correlation function $\xi(\sigma,\pi)$, where $\pi$ is the line-of-sight  or radial separation and $\sigma$ is the transverse separation. Moreover, \citet{gaz08b} have shown how to constrain $H(z)$
using the correlations in the radial direction.  \citet{kaz10a} found similar results for the correlation measurements and uncertainties, but manifest disagreement in the interpretation of the results regarding the detection of a line-of-sight baryonic acoustic feature.

More recent studies \citep{mar08a,cab08a,san09a,kaz09a} have confirmed this detection in the last Data Release (DR7, \citealp{aba08a}) of the SDSS-LRG, containing twice as many galaxies as the original sample, although the observed peak is in these cases wider than that observed  in the original detection -- an issue that needs further explanation. These measurements of the BAO scale at a low redshift, combined with other cosmological probes, have been used to put stringent constraints on the values of cosmological parameters \citep{teg06a,per07a,san09a,per10a,rei09a,kaz10a}. 

While \citet{bas10a} argue that low-level detections may not be sufficient to robustly estimate the cosmological parameters, 
\citet{cab10a} show instead that it is still possible --assuming a model-- to locate the BAO position with data providing very low significant BAO detection.

It is important, therefore, to find evidence of BAO in the galaxy distribution based on complementary methods. A step further is to search for real structures in the galaxy distribution that are responsible for the BAO feature in these second order statistics. 
The detection of these structures would be a confirmation of the existence of the baryon acoustic phenomenon. 
Moreover, if we are able to localize these structures in configuration space, this would allow us to study in more detail the properties of BAO.

In this paper, we introduce a new method for the detection of BAO, which is closely tied to the underlying physics of the process, and apply it to a sample drawn from the SDSS catalogue. This method (described in Section~\ref{sec:method}) is based on analyzing directly the 3D galaxy distribution using a very specific wavelet function (which we called `BAOlet'), which is specially well suited to search for BAO features. The method makes use of two different tracers, one to map the overall density field (including the BAO shells), and the other to locate the position of the largest overdensities, which should correspond to centres of the shells. 
As we study directly the galaxy distribution in configuration space, this method also allows us to identify regions of space where the BAO signal is stronger or fainter.
We describe the expected signal in the $\Lambda$CDM model in Section~\ref{sec:theory}, using both analytical prediction and a $N$-body simulation catalogue. We describe the samples used in the case of SDSS in Section~\ref{sec:data}. 
In Section~\ref{sec:results}, we show the results obtained in this case. 
We also make a test to assess the significance of these results, and explore the implications of this analysis regarding the localization of BAO structures.
Finally, we summarize our conclusions and discuss possibilities for future work in Section~\ref{sec:conc}.

\section{The wavelet detection method}
\label{sec:method}

The basis of the new BAO detection method is to focus on the positions of massive dark matter haloes, which correspond to the location of large initial perturbations, and to look for the presence of structures resembling the acoustic shells around these. 
Once we locate the positions of the large over-densities, we need to study the density field to identify the structures corresponding to the acoustic shells around these centres. An appropriate method for identification of structures in continuous fields is wavelet analysis \citep{mar93a,starck:book06,jon09a}. 
Wavelet transforms are widely used in many areas, especially in image analysis \citep{mallatb08,starck:book10}. They are specially suited for the  analysis of data at different scales, and identification of characteristic patterns or structures. Wavelets have been used in Cosmology for the analysis of the large-scale structure, and of the CMB anisotropies \citep{mar93a,rau93a,wave:vielva04,starck:sta05_2,saa09a}.

\begin{figure}
\begin{center}
\includegraphics[width=\columnwidth]{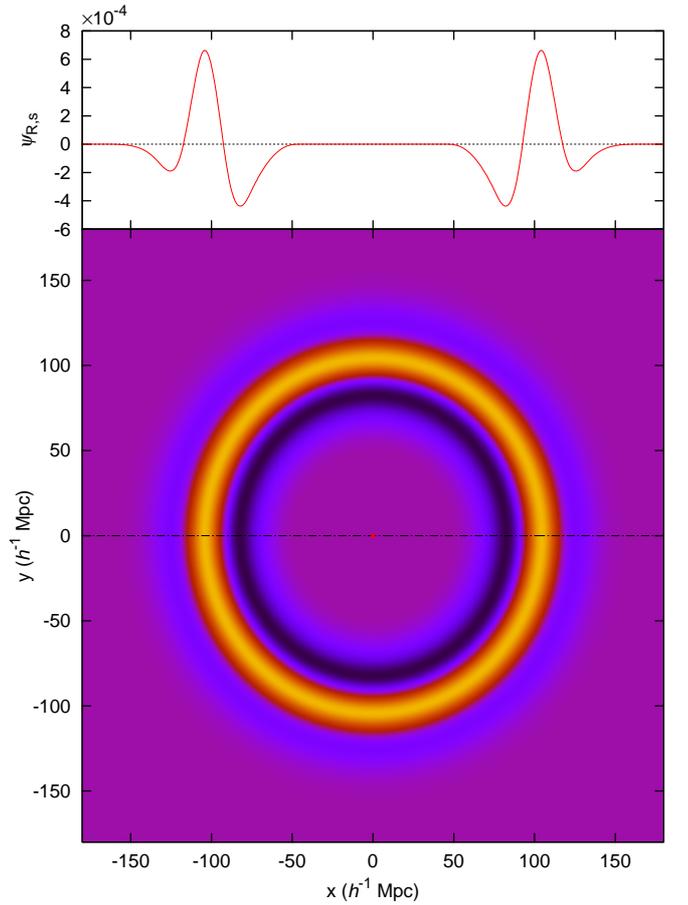}
\end{center}
\caption{The BAOlet function.
Here we show a 2D plot (bottom) of the wavelet $\psi_{R,s}(\mathbf{x})$ used in the analysis, as defined by equation~(\ref{eq:baodef}). The top panel shows
a 1D slice along the dashed-dotted axis. The wavelet is plotted here for $R = 105 \hMpc$, $s = 30 \hMpc$. 
The red dot marks the centre of the wavelet. This function has a null mean (provided that $R > 2s$), and compact support. It takes non-zero
values only for $R - 2s \leq |\mathbf{x}| \leq R + 2s$.}
\label{fig:profile}
\end{figure}

Standard wavelet functions like the Mexican hat are, however, not suitable for the detection of shells. Instead, we need a family of wavelets whose shape matches the type of structures we want to find in our data.
Therefore, we use a specially designed wavelet (the `BAOlet'), well adapted to the search of BAO features -- shell-like structures around our selected centres.
We design this new family of wavelet functions as a transformation of the wide-used B-spline wavelets \citep{saa09a}. 
 These $\psi_{R,s}(\mathbf{x})$ functions are spherically symmetric, and their radial profiles are defined as 
\begin{equation}
  \label{eq:baodef}
  \psi_{R,s}(r) = \frac{\alpha_{R,s}}{4\pi r^2} \left[ 2B_3\left( 2\frac{r - R}{s} \right) - B_3\left(\frac{r-R}{s}\right)\right] \, ,
\end{equation}
where $R$ and $s$ are the two parameters that define the scale and width of the BAOlet function, $\alpha_{R,s}$ is the normalization constant defined so that
\begin{equation}
\label{eq:wavenorm}
||\psi_{R,s}||^2 \equiv \int |\psi_{R,s}(\mathbf{x})|^2\mathrm{d}\mathbf{x} = 1 \, ,
\end{equation}
and $B_3(x)$ is the box spline of the third degree, defined by
\[
B_3(x) = \frac{1}{12} \Big(\vert x-2 \vert^3 - 4\vert x-1 \vert^3+6\vert x \vert^3-4 \vert x+1 \vert^3+\vert x+2 \vert^3 \Big) \, .
\]
The BAOlet function is shown in Fig.~\ref{fig:profile}. It can be thought of as a spherical shell of radius $R$ and width $s$, with zero amplitude at its centre and therefore adapted to the detection of spherical shells of a given radius.
This specific choice is motivated by the fact that the integrated profile  is the widely-used one-dimensional `B-spline' wavelet function that has a null mean and compact support $[-2,2]$. These properties directly translate onto the BAOlet that has also a null mean --a requirement for any wavelet function-- if $R > 2s$, and takes non-zero values only for $R - 2s \leq |\mathbf{x}| \leq R + 2s$.

We describe the density field using the density contrast $\delta(\mathbf{x})$, defined as
\[
\delta(\mathbf{x}) = \frac{\rho(\mathbf{x}) - \rho_0}{\rho_0} \, ,
\]
where $\rho(\mathbf{x})$ is the density field, and $\rho_0$ is its mean.
Then, given a density contrast map $\delta(\mathbf{x})$,
properly normalized as in equation~(\ref{eq:wavenorm}), 
we can construct, for each point in the parameter space $(R,s)$, a BAOlet coefficient map as the 
convolution of our density field with the corresponding wavelet:
\begin{equation}
  \label{eq:coeff}
  W_{R,s}(\mathbf{x}) = \int_{\Re^3} \psi_{R,s}(\mathbf{y}) \delta(\mathbf{y} - \mathbf{x}) \mathrm{d}^3\mathbf{y}\, .
\end{equation}
The BAOlet acts as a matched filter, which is sensitive to data containing shells of different radius and different widths. Its property of zero mean is also of high importance since it makes the statistics derived from the BAOlet coefficients independent of the background level. Indeed, it is obvious that any constant added to the input data would not change the BAOlet coefficients. In comparison, the estimation of such a baseline level is a very delicate aspect of the BAO detection in the two-point correlation function.
 
Due to the properties of the wavelet, the coefficient maps $W_{R,s}(\mathbf{x})$ should have a null
mean when averaged over all points in the volume considered. 
Equivalently, if we sampled these maps at $N$ random points uniformly distributed  in the volume ($\mathbf{x}_r^{(i)}$), the expected value of the average of the coefficients is zero,
\begin{equation}
\label{eq:E0}
E\left\lbrace \left \langle  W_{R,s}(\mathbf{x}_r^{(i)}) \right\rangle_N \right\rbrace = 0 \, .
\end{equation}
This condition holds even in the presence of shell-like structures in the density field.
Of course, for such structures the value of 
$W_{R,s}(\mathbf{x_c})$ ($\mathbf{x_c}$ is the centre of the shell) is positive, and remains positive in nearby points. For an ideal $\delta(r-R)$ density shell the radius of the region around the centre where the wavelet amplitude is positive, is $s$; the positive signal in this region is compensated by negative amplitudes around $|\mathbf{x}|=R$.
However, if we are able to identify the positions of $N$ massive
haloes in the same volume ($\mathbf{x}_c^{(i)}$), we can define a new statistic $B(R,s)$ as the mean value of the
coefficients $W_{R,s}(\mathbf{x})$ at these positions:
\begin{equation}
  \label{eq:Bstat}
  B(R,s) = \left\langle W_{R,s}(\mathbf{x}_c^{(i)})\right\rangle_N \, .
\end{equation}
If there are indeed shell-like structures around the selected density maxima $\mathbf{x}_c^{(i)}$, as expected for baryon acoustic structures, we should find positive values of $B(R,s)$ with the maximum of $B$ at the $(R,s)$ values characterizing these shells.

We can obtain further information from the wavelet coefficients $W_{R,s}(\mathbf{x})$, as we have information on the actual dependence of the signal picked up by the BAOlet function on the position. 
In particular, fixing a set of parameters of interest ($R_i, s_i$), we could use the coefficients $W_{R_i,s_i}(\mathbf{x}_c)$ to identify which of the selected massive haloes are giving the largest signal for these characteristics of the shells.
In the context of BAO, the parameters $R_i$, $s_i$ can be chosen \emph{a priori} using a theoretical model, or \emph{a posteriori} using the parameters for which the function $B(R,s)$ attains it maximum. 
In this way, we can localize in configuration space the structures responsible for the largest BAO signal in a given sample.

For our calculation of $B(R,s)$, we sample the $(R,s)$ parameter space on a grid. For each point $(R,s)$, we calculate the coefficient map $W_{R,s}(\mathbf{x})$ as the convolution of the BAOlet with the density field (equation~\ref{eq:coeff}). We perform the convolution in Fourier space using a Fast Fourier Transform (FFT) technique. To avoid problems with the FFT, we zero-pad a large region around our density cube. To obtain $B(R,s)$, we sample $W_{R,s}(\mathbf{x})$ at the position of the $N$ selected centres, and calculate the average value (equation~\ref{eq:Bstat}). 

Therefore, to apply this method, we need a way to map the overall density field $\delta(\mathbf{x})$, but also to locate the position of massive matter haloes $\mathbf{x}_c^{(i)}$. We have to use two different populations of mass tracers, so that they play the appropriate role in the detection algorithm. 
The idea of using two different tracer sets, one for the small perturbations and another for the high peaks, in a cross-correlation analysis was anticipated by \citet{eis07a}. We implement here a similar idea, but using a wavelet tool directly on the density field.
As detailed below, we use galaxies from the `Main' and LRG samples of SDSS in this case. However, this choice would depend on the kind of data available in each case.

\section{Prediction from $\Lambda$CDM}
\label{sec:theory}

In order to better understand our method, we show here which results 
we expect according to the $\Lambda$CDM model, and the effect of BAO in our new statistic $B(R,s)$.
We use for this aim both the analytical approximation to the transfer function of \citet{eis98a}, and the results from the MareNostrum Institut
de Ci\`encies de l'Espai (MICE) simulation \citep{fos08a}.

In the first place, we use the $\Lambda$CDM transfer function, which allow us to study directly the effect of the BAO. However, in this case,
we must do a series of approximations in order to make a prediction for $B(R,s)$. We want to predict which is the typical result for the wavelet
coefficient $W_{R,s}$ at the position of massive matter haloes $\mathbf{x}_c$, as a function of $R,s$. From equation~(\ref{eq:coeff}), we see that this 
is equivalent to study the typical density profile around such haloes, $\delta(\mathbf{y}-\mathbf{x}_c)$. The $\Lambda$CDM transfer function
allows us to calculate
this profile, provided we know which is the initial perturbation corresponding to the selected haloes. 
We make here the simple approximation of  considering that these initial perturbations are point-like and spherically symmetric, and can thus be simply described by a Dirac delta function in configuration space.
This corresponds to a constant value in Fourier space. 
As the transfer function $T(k)$ describes the relative evolution of the different Fourier modes, 
the present day radial density profile corresponding to such initial perturbation will we given simply by \citep{eis07a}
\begin{equation}
  \label{eq:tf}
  \rho(r) = C \widetilde{T}(r) \,
\end{equation}
where $\widetilde{T}(r)$ is the Fourier transform of the transfer function $T(k)$, and $C$ is a normalization constant that depends on the details
of the initial perturbation, and on the cosmic growth function $D_1(z)$. From equations~(\ref{eq:coeff}) and (\ref{eq:Bstat}), we see that the effect
of $C$ will be just to change the normalization of our statistic $B(R,s)$.

\begin{figure}
\begin{center}
\includegraphics[width=\columnwidth]{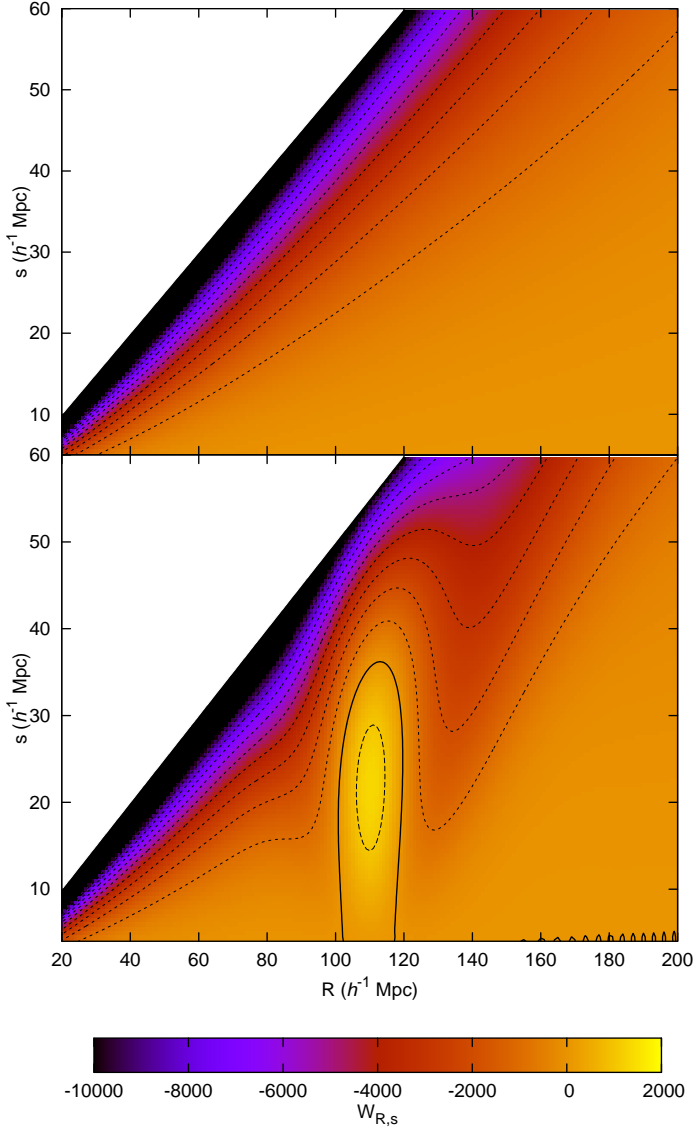}
\end{center}
\caption{Values of the BAOlet coefficients $W_{R,s}$ expected at the positions of large initial
point-like perturbations, as a function of the BAOlet parameters $(R,s)$. The bottom panel shows the
result using a standard $\Lambda$CDM transfer function, while the top panel shows the result using a 
transfer function with the BAO wiggles edited out \citep{eis98a}. The normalization is arbitrary.
The contours are drawn at steps of 1000 for $W_{R,s}<0$ (dotted), $W_{R,s}=0$ (solid), and $W_{R,s}>0$ (dashed). 
The map attains a maximum at $R = 110 \hMpc$, $s = 22 \hMpc$.
}
\label{fig:eishu}
\end{figure}

We used the fitting formulae to the transfer function $T(k)$ from \citet{eis98a}, and obtained the expected $W_{R,s}$ at a large overdensity using equations~(\ref{eq:tf}) and (\ref{eq:coeff}). In order to highlight the particular signature of BAO, we also calculated $W_{R,s}$ using the `no wiggle' transfer function formula, in which the BAO have been edited out. We used here the values $\Omega_M  =0.25$, $\Omega_{\Lambda} = 0.75$, $\Omega_b = 0.044$, and $h=0.7$ for the cosmological parameters, to allow a direct comparison with the MICE simulation. Following \citet{eis98a}, the sound horizon scale in this case is $r_s = 109.3 \hMpc$. 
The results for both cases are shown in Fig.~\ref{fig:eishu}. In the plot, we mask the region
$R < 2s$, as for these values of the parameters our BAOlet is not compensated (its mean is different from $0$). 
Comparing both panels of the Figure, we see clearly which is the effect of the presence of BAO in our statistic. 
In the case without BAO $W_{R,s}$ is always negative, and it presents a smooth gradient across the $(R,s)$ plane. This
gradient is due to the overall shape of the radial profile (equation~(\ref{eq:tf}).
However, in the presence of BAO, $W_{R,s}$ shows a prominent peak with positive values. This clearly shows the idea behind the $B(R,s)$ statistic. The BAOlet $\psi_{R,s}$ acts as matched filter with a shape adapted to detect BAO shells. Therefore the positive values in the coefficients $W_{R,s}$ correspond to the cases in which the radial profile is matched by the BAOlet shape. The values at which $W_{R,s}$ attains its absolute maximum, $R_{\mathrm{max}} = 110 \hMpc$ and $s_{\mathrm{max}} = 22 \hMpc$, correspond thus to the characteristics of the shell that best matches the observed profile about the selected centres.

In order to test the reliability of the method, and of this $\Lambda$CDM prediction, we calculated the $B(R,s)$ for a halo catalogue drawn from the MICE simulations. We used the publicly available halo catalogue from the `MICE3072' run \citep{cro10a}. This particular run
contains $2048^3$ particles in a box of side $3072 \hMpc$, therefore covering a volume of $29 \, h^{-3} \, \mathrm{Gpc}^3$. The simulation was run with 
the GADGET-2 code \citep{spr05b}, assuming a $\Lambda$CDM model with the parameters mentioned above. Haloes in the simulation were selected using a friends-of-friends (FoF) algorithm.

We used the resulting halo catalogue at $z=0$, which contains a total of $2819031$ haloes containing $143$ or more particles. This corresponds to haloes
with masses $\geq 3.35 \times 10^{13} \hMs$. The halo number density is thus $9.72 \times 10^{-5} \, h^3 \mathrm{Mpc}^{-3}$. We used the full halo catalogue
as a tracer of the overall density field. We then selected as centres for the calculation of $B(R,s)$ in equation~(\ref{eq:Bstat}) only the haloes
with a mass $\geq 1.76 \times 10^{14} \hMs$. We chose this mass threshold in order to select approximately the $10\%$ most massive haloes in the simulation
box. This choice is somewhat arbitrary, but serves for the purpose of testing the BAOlet method and illustrating the expected result. 

\begin{figure}
\begin{center}
\includegraphics[width=\columnwidth]{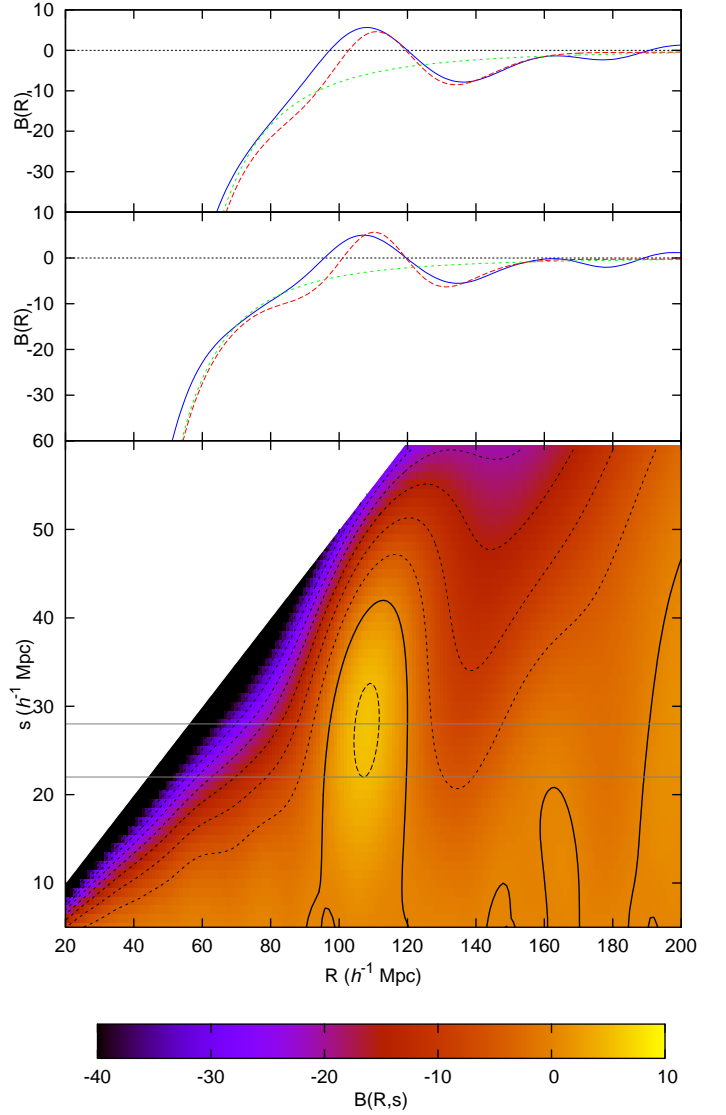}
\end{center}
\caption{The BAOlet statistic $B$ calculated for the MICE simulation sample described in the text as a function of the parameters $(R,s)$ (bottom panel).
The contours are drawn at steps of $5$ for $B<0$ (dotted), $B=0$ (solid), and $B>0$ (dashed). This function attains its maximum for $R=108 \hMpc$, $s=28 \hMpc$.
The top two panels show cuts at the values $s = 28 \hMpc$ (top)
and $s = 22 \hMpc$ (middle), marked with grey horizontal lines in the 2D panel.
In each case, the solid blue line corresponds to the value obtained from MICE, the dashed red line
corresponds to the theoretical expectation from the \citet{eis98a} transfer function (bottom panel of Fig.~\ref{fig:eishu}), 
and the dotted green line to the theoretical expectation using the `no wiggle' transfer function (top panel of Fig.~\ref{fig:eishu}).
These theoretical predictions have been re-normalised to get the same value at the maximum in $B(R,s)$.
}
\label{fig:mice}
\end{figure}

Fig.~\ref{fig:mice} shows the BAOlet result $B(R,s)$ for these MICE samples, compared to the theoretical results obtained above from the \citet{eis98a} transfer functions. We obtain a result very similar to that of Fig.~\ref{fig:eishu}, as $B(R,s)$ shows a clear peak, and attains its absolute maximum for 
$R_{\mathrm{max}} = 108 \hMpc$, $s_{\mathrm{max}} = 28 \hMpc$. This indicates that our BAOlet method can be applied to two sets of mass tracers, although the details of the tracers used here
are very different from the ones we use later on the SDSS samples. This also confirms the expected effect of the presence of BAO in the $B(R,s)$ function:
the presence of a large peak with positive values of $B$, located approximately at the values of $R$ and $s$ corresponding to the radius and width of the
acoustic shells. The fact that we obtain here slightly different values for $R_{\mathrm{max}}$ and $s_{\mathrm{max}}$ than those predicted above may be due to 
non-linear evolution effects, which slightly reduce the radius and increase the width of the shells. A similar effect is present in the correlation 
function \citep[see e.g.][]{cro08a}.

\begin{figure}
\begin{center}
\includegraphics[width=\columnwidth]{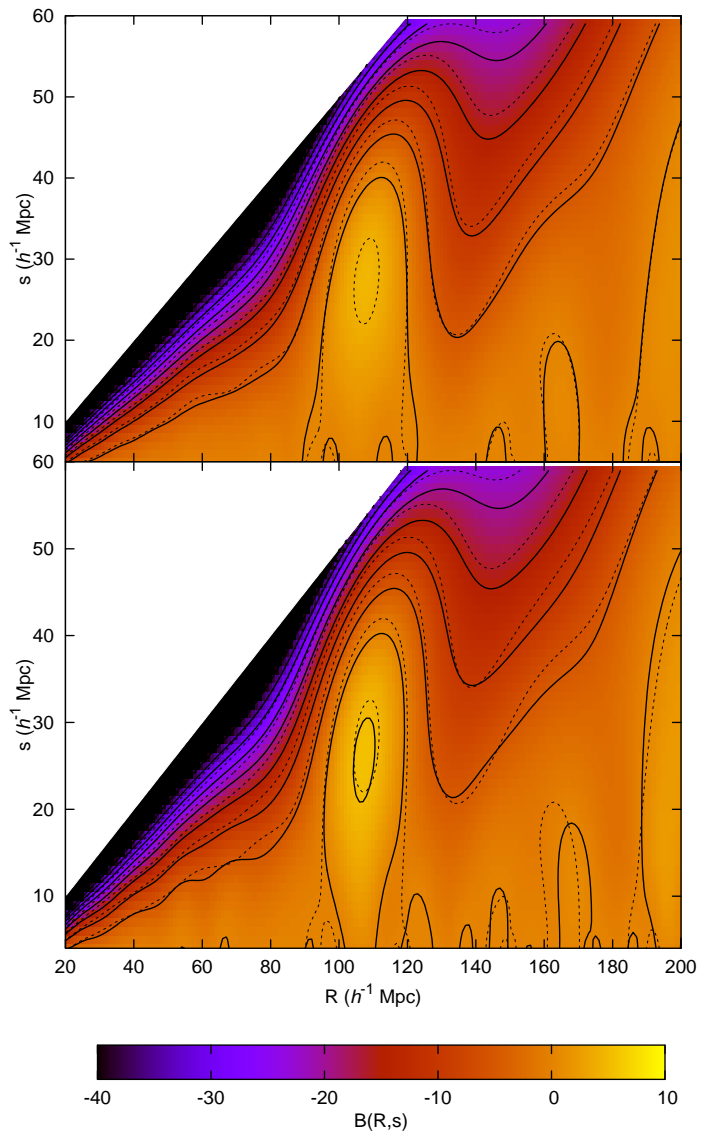}
\end{center}
\caption{
The BAOlet statistic $B$ for the MICE simulation when some observational effects are taken into account.
In the top panel, we show the $B(R,s)$ obtained when redshift-space distortions are included in the simulation.
In the bottom panel, we show the result when a radial selection function is applied to the halo catalogue.
In both cases, the contours are drawn at steps of $5$ in $B$. 
Solid contours correspond to the results with the observational effects included.
The dashed contours correspond to the original real-space result without selection, i.e., they are identical to those in the bottom panel of Fig.~\ref{fig:mice}.
}
\label{fig:mice-obs}
\end{figure}

We also used this halo catalogue from MICE to make a qualitative estimation of how different observational effects would affect the BAOlet result.
In the first place, we studied the effect of redshift-space distortions. 
To this end, we calculated the redshift-space positions of all haloes taking into account their peculiar velocities, as output by the simulation, and considering an observer located in one of the vertices of the simulation cube. 
The result for $B(R,s)$ in this case is shown in the top panel of Fig.~\ref{fig:mice-obs}, where it is compared to the real-space result discussed above. 
As can be seen from the figure, although small differences appear between the real- and redshift-space results, the main features of the $B(R,s)$ prediction remain the same, with the position of the maximum changing only by $\sim 1 \hMpc$.

For the second case, we added the effect of a decreasing radial selection function across the sample. 
We model this selection as an exponential decay function, such that the final number of haloes used to trace the overall density field is $\sim 20\%$ of the total. 
In our calculations, we then weight each halo by the inverse of the mean density at its redshift, as we do later for the SDSS data. 
We do not apply any selection function to the centres. 
The results for $B(R,s)$ obtained in this case (including also the redshift-space effects) are shown in the bottom panel of Fig.~\ref{fig:mice-obs}. 
As above, these observational effects do not change significantly the overall behaviour of $B(R,s)$, or the location of the maximum of the peak.
Overall, although the MICE catalogue used does not mimic the characteristics of our SDSS samples, we can be confident that neither redshift-space distortions nor a radial selection function (when it is taken into account in the calculation) should bias significantly our results.

\section{SDSS samples used}
\label{sec:data}

We used data from two different samples of the latest data release (DR7) of the spectroscopic SDSS.  On one side, we used the `Main' galaxy sample \citep{str02a} as mass tracers for reconstructing the overall density field $\delta(\mathbf{x})$. On the other, we used the LRGs as tracers of the central over-densities, and therefore used them as the selected centres $\mathbf{x}_c^{(i)}$ to compute $B(R,s)$.

Luminous Red Galaxies were selected by the SDSS team using several colour and magnitude cuts to obtain a highly biased sample reaching high redshift \citep{eis01a}. The galaxies selected in this way are known to reside near the centres of massive dark matter haloes \citep{zhe09a} and are thus adequate tracers for the centres of baryon acoustic structures. We applied an extra cut in the K-corrected, evolved, $g$-band absolute magnitude: $-23.2 < M_g < -21.2$, as in the previous BAO analysis by \citet{eis05a}. This results in an approximately volume-limited sample in the redshift range $0.15 < z < 0.30$. 

`Main' galaxies in the SDSS constitute a much denser sample, and are therefore more suitable to map small density changes such as BAO shells. We used the `Main' sample from the Value-Added Galaxy Catalogue \citep{bla05b}, which constitutes a magnitude limited sample in the $r$ band, with $r < 17.6$. We applied an extra simple cut, $M_r < -20$.

For the conversion of angles and redshifts into co-moving distances, we used a fiducial cosmology with the parameters $\Omega_M = 0.25$, $\Omega_{\Lambda} = 0.75$. 
In all our analysis we use distances in units of $\hMpc$, so that they do not depend on the specific value of $h$.
We converted the distribution of the `Main' galaxies into a density field $\delta(\mathbf{x})$ binning it into a grid with cubic pixels of $3 \hMpc$ side. 
We corrected for the selection effects by weighting each galaxy by the inverse of the average density at its redshift.
As explained below, we performed some tests by slightly changing this weighting scheme.
Although this weighting may not be optimal, it should not affect significantly our results, given that the wavelet method does not depend on the local background level.
 We used the density field constructed in this way for the calculation of the BAOlet coefficients following equation~(\ref{eq:coeff}).

In our calculations, we could only use the region in which these two samples overlap, which corresponds to the redshift limits $0.15 < z < 0.26$. 
To minimize border effects in the $B(R,s)$ calculation, we defined a buffer region of $r_{\rm buff} = 175 \hMpc$ from any of the borders of the `Main' sample volume. We used as centres only the LRGs in the inner volume. This allows us to use the density field, as traced by the `Main' sample galaxies, from $z > 0.09$.
In order to minimize angular selection effects and border effects, we use a compact area of the sky where the angular completeness is nearly uniform. This area covers $5511 \deg^2$ and is defined, in the SDSS survey coordinates \citep{sto02a}, by the limits $-31.25^{\circ} < \eta < 28.75^{\circ}$, $-54.8^{\circ} < \lambda < 51.8^{\circ}$. This results in finally using the density field in a volume of $2.2 \times 10^8\,h^{-3}\,\mathrm{Mpc}^3$, as traced by $N_{\rm Main} = 198342$ galaxies. 
The number of LRGs used as centres (avoiding the buffer region) is $N_{\rm LRG} = 1599$.

\begin{figure*}
\begin{center}
\includegraphics*[width=0.8\textwidth]{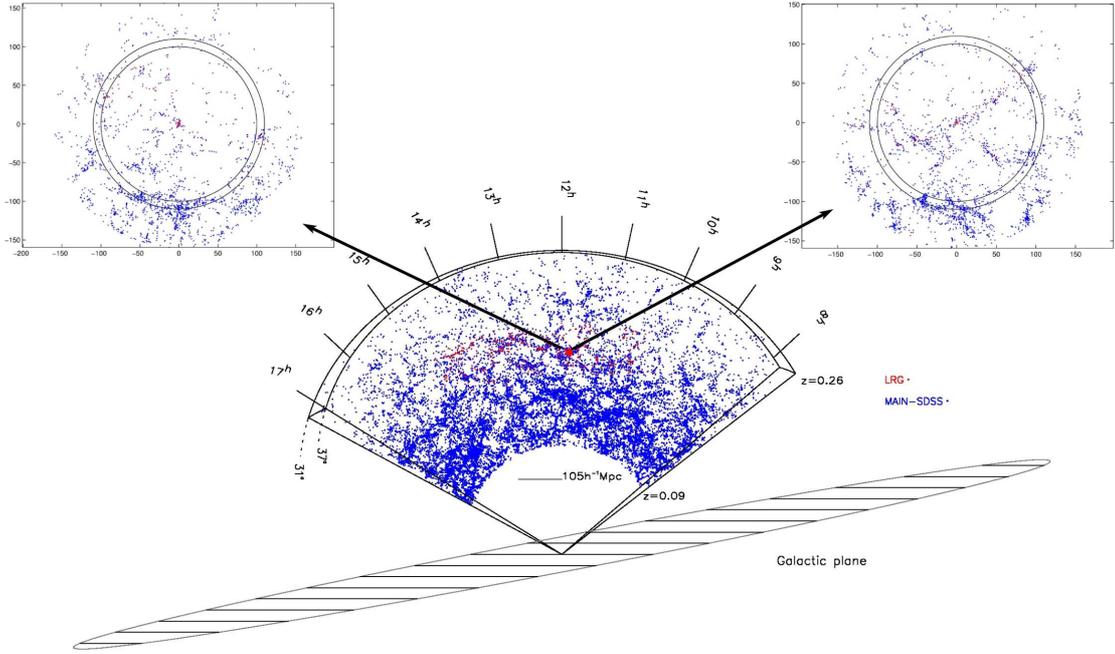}
\end{center}
\caption{
The SDSS catalogues used.  
We show a 6 degree thick slice drawn from the catalogues used in our analysis. 
The red points are the LRG galaxies, which form a nearly volume-limited sample at $z>0.15$. 
The `Main' galaxies of the SDSS are depicted in blue. 
As shown, we use the `Main' galaxies from $z>0.09$, and we use only the LRGs in an inner volume, allowing for a buffer region of $r_{\rm buff} = 175 \hMpc$ from any of the borders of the `Main' sample volume.
For our analysis, we use the samples covering a total area of $5511 \deg^2$. 
The radius of a typical BAO shell is shown as a segment. 
At the top insets we show two orthogonal slices of width $20 \hMpc$,  centred on a particular LRG. 
The BAOlet coefficient map $W_{R,s}(\mathbf{x})$  has a large value at the position of this centre (for $R_{\rm max}, s_{\rm max}$), and thus we expect to find a strong BAO signal. 
The two circles have the radii of 100 and 110 $\hMpc$ respectively. 
This is a single BAO shell where the over-density can be appreciated by eye at the right scale.
}
\label{fig:slices} 
\end{figure*}

In Fig.~\ref{fig:slices} we show a slice of this survey showing both the `Main' galaxies and the LRGs. 
We show how, given the buffer used, the LRGs used as centres are located only in an inner volume of the larger `Main' sample.
To illustrate the idea of the method, we show a zoom around a given LRG galaxy. Even for this single centre, a slight over-density of `Main' galaxies is seen at the radii of 105--110$\,\hMpc$. 

As the structures we look for are huge, with radii about $100 \hMpc$, we have to consider the effect of assumed cosmology (different comoving distances) on our result. 
In order to estimate the distance differences, we compared the distances in our adopted MICE cosmology ($\Omega_M = 0.25$, $\Omega_{\Lambda}=0.75$) with these in the WMAP 7-year cosmological model \citep{kom11a}, $\Omega_M = 0.271$, $\Omega_{\Lambda}=0.729$. 
We fixed the redshift difference $\delta z=0.07$ that corresponds approximately to our shell diameter of $200 \hMpc$, and found that this gives distance differences of only a 0.3 and 0.8 per cent at the near and far borders of our sample (the MICE distances are larger than the WMAP7 ones in each case).  
So, for our nearby volume, the effect is small, and does not affect our results given that the statistical uncertainties are much larger (see next Section). However, this effect will be significant for deep samples.

\section{Results for the SDSS samples}
\label{sec:results}

\begin{figure}
\begin{center}
\includegraphics[width=\columnwidth]{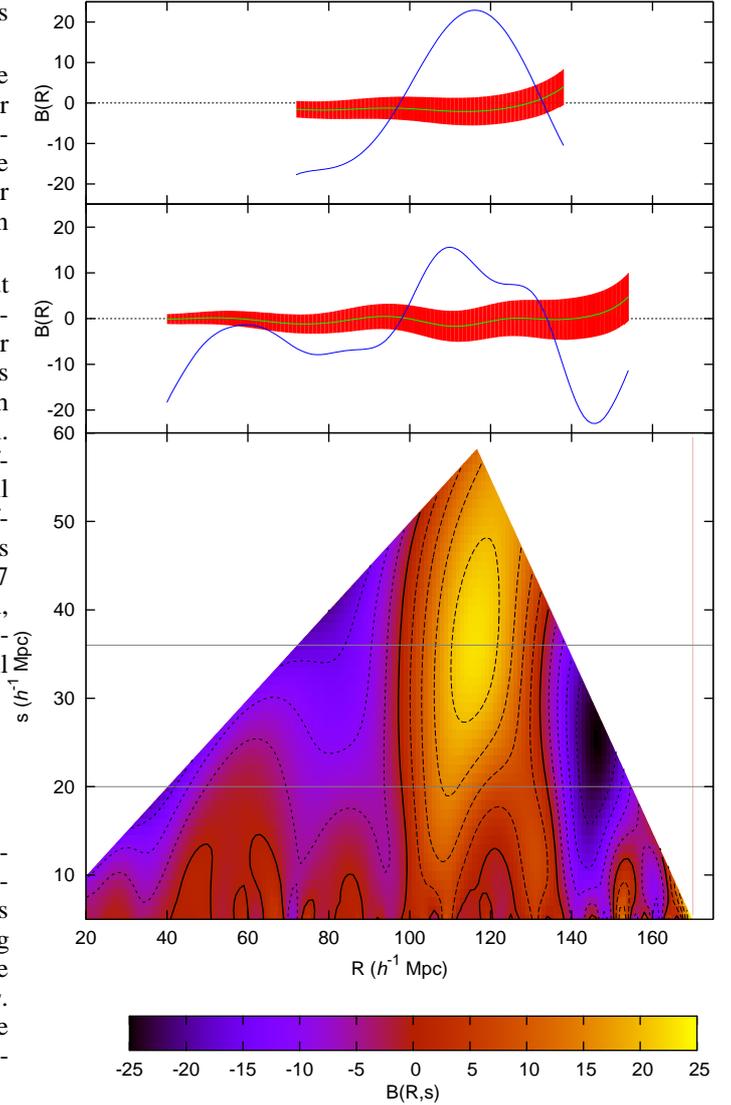}
\end{center}
\caption{The BAOlet statistic $B$ calculated for SDSS data as a function of the parameters $(R,s)$. 
The bottom panel shows the results in the full parameter space considered, where we sampled both $R$ and $s$ at intervals of $1\hMpc$. We mask two areas, at the upper right and left corners, where our results are not reliable (see details in the text). 
The contours are drawn at steps of 5 for $B<0$ (dotted), $B=0$ (solid), and $B>0$ (dashed). 
The top two panels show cuts at the arbitrarily chosen values $s = 36 \hMpc$ (top) and $s = 20 \hMpc$ (middle), marked with grey horizontal lines in the 2D panel. In these panels, the blue line is $B(R,s)$, while the green line and the red band show the mean ($\overline{B}^{MC}$) and 1-$\sigma^{MC}$ interval for the Monte Carlo realizations of random centres. We obtain a clear significant peak at different values of $s$, with a maximum for $R = 116 \hMpc$, $s = 36 \hMpc$.
}
\label{fig:baolet}
\end{figure}

We performed the calculation of $B(R,s)$ for the SDSS in an analogous way to the case of the MICE simulation, using the samples defined in Section~\ref{sec:data}. Our results are shown in Fig.~\ref{fig:baolet}. As above, we mask the region $R < 2s$. 
As we are not introducing any border correction when calculating the $B(R,s)$ statistic, we also mask the region corresponding to the values $R > r_{\rm buf} - s$. Values obtained at a those large values of $R$ could contain some spurious signal, as the calculation of $W_{R,s}$
would rely on the density field in regions outside of the survey boundaries.

The resulting $B(R,s)$ map is qualitatively very similar to that expected, either using an analytical $\Lambda$CDM model (Fig.~\ref{fig:eishu}), or the MICE simulation (Figs.~\ref{fig:mice} and \ref{fig:mice-obs}). 
This is an indication that the observed pattern does not originate from spurious features in the SDSS but is closely related to the large scale structure and more specifically the BAO. 
$B(R,s)$ attains a maximum at $R_{\rm max} = 116 \hMpc$, $s_{\rm max} = 36 \hMpc$. 
This maximum is clearly related to the characteristics of the BAO structures present in our samples. 
We studied the robustness of this result by changing the weighting scheme applied for the construction of the density map (see Section~\ref{sec:data}).
We did so by capping at different maximum values the possible weights associated to each galaxy, and repeating the calculation of $B(R,s)$ in each case.
The results were qualitatively similar, obtaining a peak in $B(R,s)$ in all cases. 
However, the position of the peak changed in each case, with maximum changes of the order of $\pm 5 \hMpc$ in $R_{\rm max}$, and $\pm 10 \hMpc$ in $s_{\rm max}$.
Therefore, the difference between the position of the peak obtained from the SDSS data and that given by the MICE simulation is not significant.
In any case, we can not use the scale and the width of the observed maximum of $B(R,s)$ as direct estimates of the radius or width of the shells, specially given that our analysis of the possible observational biases (Fig.~\ref{fig:mice-obs}) was only qualitative.

In order to assess the significance of the BAO detection with this method, we focused on the value of $B(R,s)$ obtained at the maximum, $B_{\rm max} = B(R_{\rm max}, s_{\rm max}) = 22.9 \pm 3.7$\footnote{This error in $B_{\rm max}$ is obtained from the variance of the coefficients at the $N_{\rm LRG}$ different LRGs. However, our significance test is independent of this error value.}. A more thoroughfull analysis would model the $B(R,s)$ statistic in the full parameter space. However, given the large covariances between measurements at different values of $(R,s)$ we do not expect a large difference from the simple case we consider. We will assess the probability of finding such a maximum in the case in which there are not baryon acoustic structures present in our sample. We model this null hypothesis by using randomly distributed centres for the calculation of $B(R,s)$ in equation~(\ref{eq:Bstat}), instead of LRGs. Even using the $W_{R,s}(\mathbf{x})$ coefficients from the observed density field (traced by SDSS `Main' galaxies), the expected value of $B(R,s)$ in this case is 0 (see equation~\ref{eq:E0}), and we expect to obtain a significantly higher signal in the data. In this way, we are testing the null hypothesis that, either there are not shell-like structures in the density field traced by the `Main' sample, or these shell-like structures are not found preferentially around LRG centres. In either case, that would mean that there are not BAO-like structures present in our sample.

To perform the significance test, we generated $10^5$ random realizations of a Poisson process, with the mean number of points $N_{\rm LRG}$, in the same volume as the LRGs considered in the calculation (i.e. taking into account the buffer zone). For a realization $j$, we use the generated points as our centres $\mathbf{x}_c^{(i)}$ to compute the $B(R,s)$ statistic following equation~(\ref{eq:Bstat}), using the $W_{R,s}(\mathbf{x})$ coefficients obtained from the data. We can then obtain the mean value 
$\overline{B}^{MC}(R,s)$, and the standard deviation $\sigma^{MC}(R,s)$ of the Monte-Carlo realizations of the centres. We show $\overline{B}^{MC}(R,s)$ and a band of $1 \sigma^{MC}(R,s)$ around it in the top panels of Fig.~\ref{fig:baolet}. 

\begin{figure}
\begin{center}
\includegraphics[width=\columnwidth]{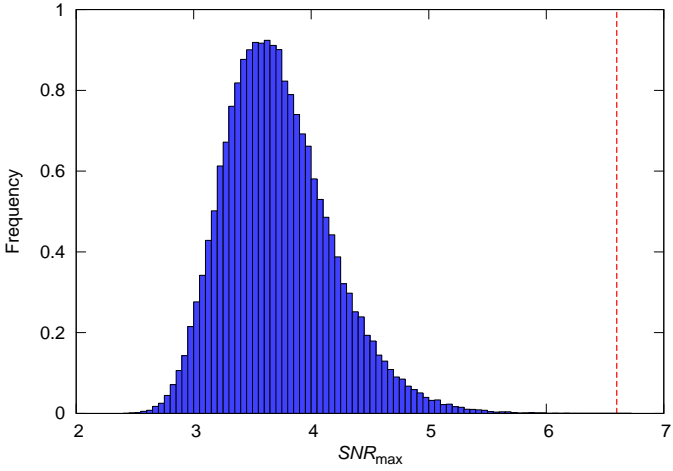}
\end{center}
\caption{Histogram showing the distribution of the maximum $SNR$ values obtained, in the full $(R,s)$ space, for the $10^5$ Monte Carlo realizations of Poisson-distributed centres ($SNR^{MC(j)}_{\rm max}$). This histogram has a mean of $3.72$ and a standard deviation of $0.46$. We show as a dashed vertical line the value obtained from the data (using the LRGs as centres), $SNR_{\rm max} = 6.60$. Only one of the Monte Carlo realizations give a maximum value larger than $SNR_{\rm max}$.}
\label{fig:histogram}
\end{figure}

We now calculate our signal-to-noise ratio at the maximum as $SNR_{\rm max} \equiv B_{\rm max}/\left[\sigma^{MC}(R_{\rm max},s_{\rm max})\right] = 6.60$, and assess the probability of finding such a large value of $SNR_{\rm max}$ anywhere in the parameter space for the Monte Carlo realizations. We used $SNR_{\rm max}$ instead of directly using $B_{\rm max}$ because for some regions of parameter space, specially at low $s$, $\sigma^{MC}(R,s)$ is extremely large. Therefore, if we used $B_{\rm max}$, we would need to arbitrarily restrict the parameter space studied, thus introducing a possible a posteriori bias. When using $SNR_{\rm max}$ we sample the full parameter space considered in the calculations (as shown in Fig.~\ref{fig:baolet}). We computed the maximum value of $SNR$ for each realization $j$ in the full $(R,s)$ range, $SNR^{MC(j)}_{\rm max}$. The distribution of the values of $SNR^{MC(j)}_{\rm max}$ is shown in Fig.~\ref{fig:histogram}, where it is compared to the value of $SNR_{\rm max}$ obtained in the real data. We found that only one of the realizations gave a value of $SNR^{MC(j)}_{\rm max}$ larger than $SNR_{\rm max}$. Thus, the probability of obtaining a maximum with such a large $SNR$ in the absence of baryon acoustic structures (our null hypothesis) is $p \simeq 10^{-5}$, equivalent to a $\sim 4.4\sigma$  detection in the Gaussian case.

However, we should stress here that the significance found in this work can not be compared directly to other detection levels 
found in the literature, as it has been stated in the introduction. 
In particular, we are not comparing our results with an analytical no-BAO model of $B(R,s)$ (such as that shown in the top panel of Fig.~\ref{fig:eishu}), since to do so would require the detailed modelling of all the selection effects affecting the two samples used. 

As explained in Section~\ref{sec:method}, we can extract more information about the BAO phenomenon in our samples making further use of the BAOlet coefficient maps $W_{R,s}(\mathbf{x})$. 
Here, we use the coefficient values at the positions of the LRG, for the parameters $R_{\rm max}$, $s_{\rm max}$, which correspond to the characteristics of the BAO shells present in our samples.
In this way, the values $W_{\rm max} \equiv W_{R_{\rm max}, s_{\rm max}}$ are a measure of how strong is the signal coming from a BAO shell around a given point, and in particular, a given LRG.
Therefore, using $W_{\rm max}$ we can localise in configuration space the regions of the volume covered by our samples where the BAO signal is mostly coming from.

\begin{figure*}
\begin{center}
\includegraphics*[width=0.8\textwidth]{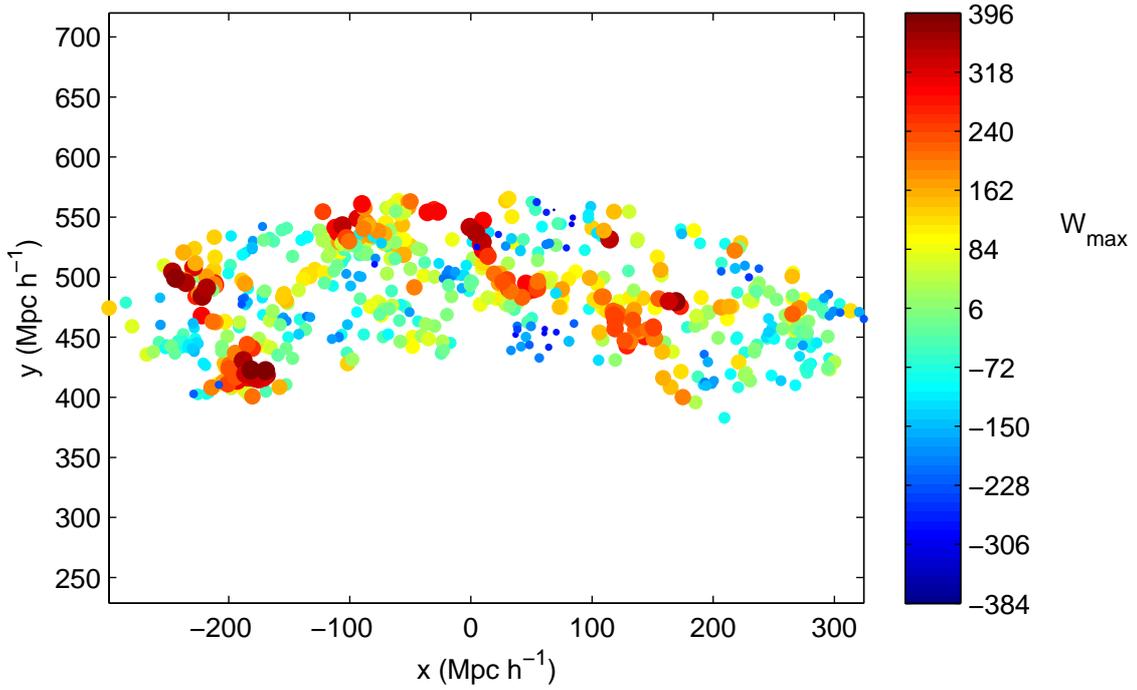}

\end{center}
\caption{A two-dimensional projection of the distribution in redshift space of the LRGs used in the analysis (i.e. inside our buffer region). 
The data showed corresponds to a slice of a width of $\simeq8^{\circ}$ in declination, which contains $\sim 40\%$ of the LRGs.
The colour and size of each point corresponds to the value of $W_{\rm max}$ at this LRG position. 
}
\label{fig:3dcoeff} 
\end{figure*}

 We illustrate this idea in Fig.~\ref{fig:3dcoeff}, where we plot a two-dimensional projection of the distribution of the LRGs used as centers in our analysis, showing also the value of $W_{\rm max}$ for each of them, following a color gradient.
The highest values of $W_{\rm max}$ correspond to the red points in the plot.

In Table~\ref{tab:gallist}, we provide the 10 LRGs used as centers with the larger values of  $W_{\rm max}$. The whole  catalogue of the $N_{\rm LRG} = 1599$ LRGs used as our centers, and the value of $W_{\rm max}$ obtained for each of them can be found
at the web page {\tt http://www.uv.es/martinez}. 
This catalogue could be used to study the relation of the BAO signal at a given LRG to its properties or the environment. 
It could also be used to make a selection of LRG centres with high signal, and use them to refine the measurements of the BAO characteristics.

\begin{table*}
  \centering
\begin{tabular}{lccccc}\hline
SDSS object name & $\alpha$ (deg) & $\delta$ (deg) & $z$ & $\sigma_z$ & $W_{\rm max}$ \\ \hline
SDSS J141746.20+184733.0 & 214.44254 &  18.79250 &  0.19872 & 0.00016 &  517.84 \\ 
SDSS J121858.41+380813.6 & 184.74341 &  38.13714 &  0.18974 & 0.00018 &  436.68 \\ 
SDSS J112430.27+415557.3 & 171.12613 &  41.93260 &  0.19433 & 0.00019 &  419.45 \\ 
SDSS J112355.53+423816.5 & 170.98140 &  42.63793 &  0.19404 & 0.00020 &  414.63 \\ 
SDSS J112352.72+424542.4 & 170.96968 &  42.76178 &  0.19469 & 0.00018 &  414.63 \\ 
SDSS J122935.13+384636.4 & 187.39640 &  38.77680 &  0.18686 & 0.00016 &  413.47 \\ 
SDSS J112535.99+412608.3 & 171.39998 &  41.43564 &  0.19288 & 0.00019 &  401.91 \\ 
SDSS J104501.94+362944.3 & 161.25810 &  36.49566 &  0.15938 & 0.00015 &  399.06 \\ 
SDSS J140443.31+264439.2 & 211.18047 &  26.74424 &  0.15854 & 0.00016 &  396.19 \\ 
SDSS J142031.28+211700.4 & 215.13036 &  21.28346 &  0.19232 & 0.00020 &  395.40 \\ \hline

\end{tabular}
\caption{Table containing the basic characteristics (J2000 sky coordinates, redshift and redshift uncertainty) of the 10 LRGs with
the larger values of  the BAOlet coefficient at the maximum $W_{\rm max}$.
 The full table is available at {\tt http://www.uv.es/martinez}.}
  \label{tab:gallist}
\end{table*}

\begin{figure}
\begin{center}
\includegraphics[scale=0.45]{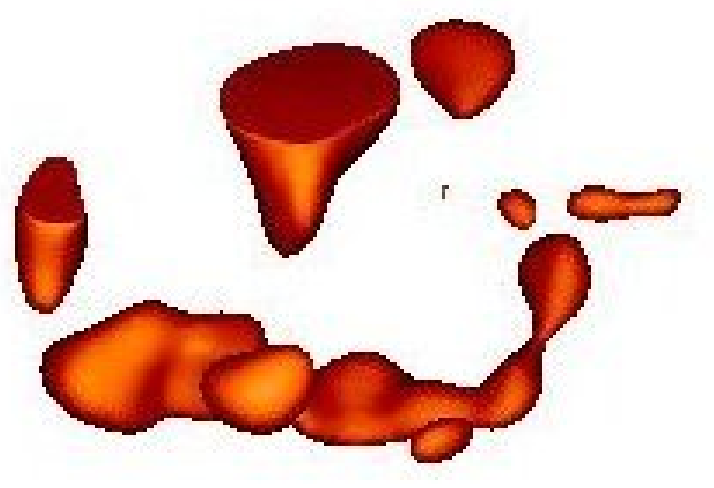}\\
\includegraphics[scale=0.45]{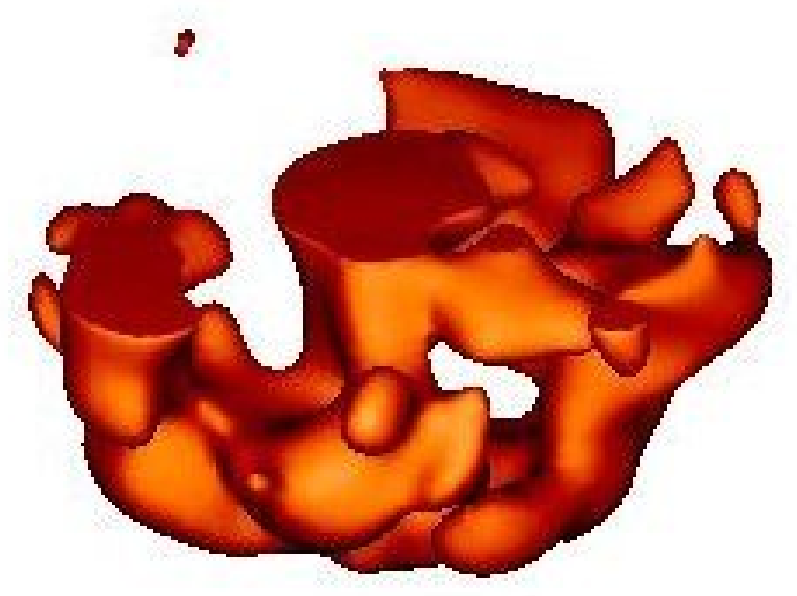}\\
\includegraphics[scale=0.45]{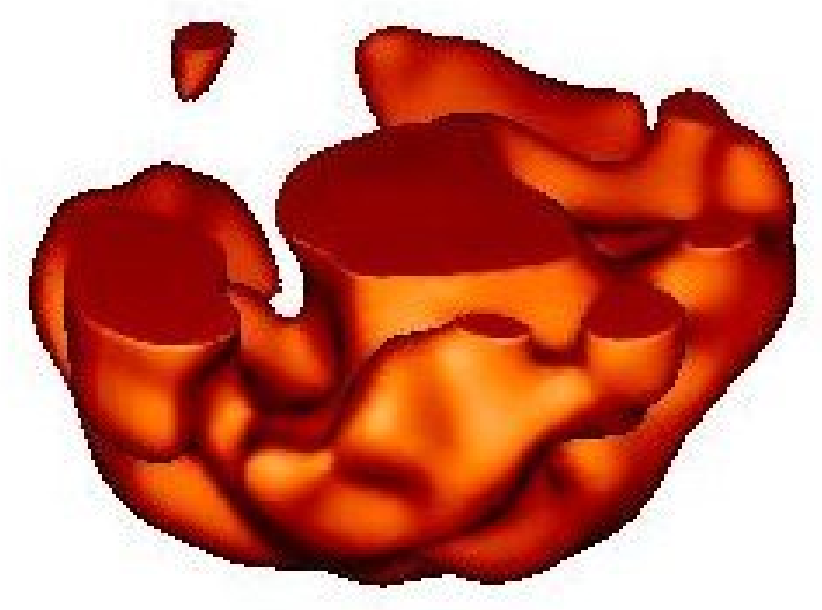}\\
\includegraphics[scale=0.60]{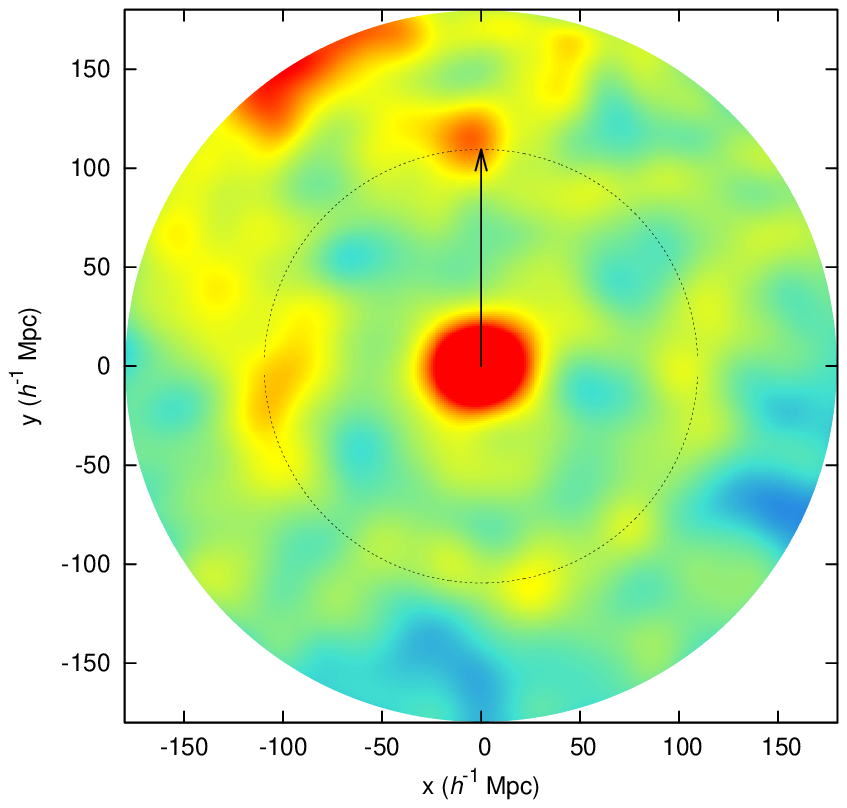}
\end{center}
\caption[]{Stacked 3D and 2D density field. 
In the top three panels, we show the density field after stacking the $N_r$ centres with $W_{R_{\rm max}, s_{\rm max}} > 0$.  
We show surfaces encompassing the regions above different thresholds in density after an isotropic Gaussian smoothing with $\sigma = 10 \hMpc$. 
The density threshold decreases from top to bottom, with values of $\delta = 1.24, 1.18, 1.13$ ($\delta$ is the density relative to the average density of the sample). 
We show only the bottom half of the density field for clarity. 
It can be seen that the acoustic shell appears clearly around the central over-density at the detected horizon scale. 
A 2D slice of this density field is shown in the bottom panel.
Here, the dotted line is a circle whose radius corresponds to the one we measure for the BAO shells, $r_{\rm max} = 109.5 \hMpc$, and the arrow marks the direction of the line of sight.
}
\label{fig:stack3d}
\end{figure}

 As an illustration of this later use, we show a simple way to study the overall properties of the BAO structures, its shape and scale. 
We select those centres which we know that present a prominent acoustic feature, i.e., those for which $W_{\rm max}> 0$
This leaves us with $N_r = 809$ centres. In order to improve the signal-to-noise in this illustration for studying the BAO structures, we stacked together the 3D density maps around the $N_r$ selected LRGs. In doing so, we kept the line-of-sight direction aligned for all the centres, as this direction will define the possible anisotropies in the distribution. We show a 3D view and a 2D cut of this stacked density map in Fig.~\ref{fig:stack3d}. Thanks to this selection the characteristic elements of the BAO are amplified: on the one side, a central bump with high density, corresponding to the massive halo traced by the LRG, and on the other side, the shell surrounding it at a scale of $\sim 109 \hMpc$, showing a fainter over-density. We also observe the anisotropic nature of these structures. This is a combination of the fact that we have to work in redshift space, and of the redshift-dependent selection function for `Main' galaxies.

\begin{figure}
\begin{center}
\includegraphics[width=\columnwidth]{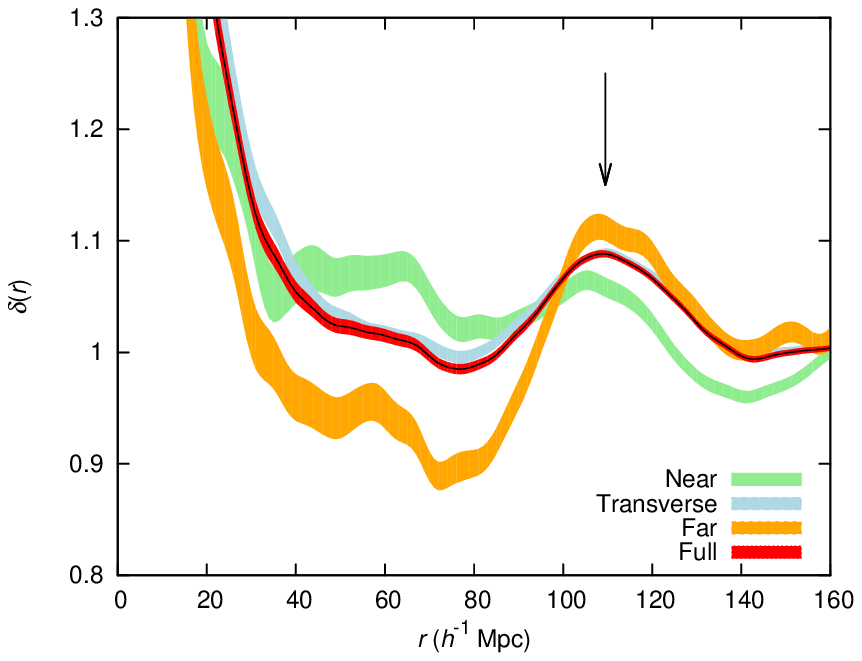}
\end{center}
\caption[]{Average radial profiles.
We show the radial density profile averaged over the $N_r$ centres with $W_{R_{\rm max}, s_{\rm max}} > 0$. 
We plot $\delta(r) = \frac{\rho(r)}{\rho_0}$, where $\rho_0$ is the average density of the sample. 
The continuous line with red error band shows the radial profile for the full sphere. 
We also show the error bands for the radial profile restricted to regions of the sphere, as defined in the text: the `near' region (green), the `far' region (blue), and the `transverse' region (orange). 
The arrow signals the location of the maximum, $r_{\rm max} = 109.5 \hMpc$. 
The error band corresponds to the 1-$\sigma$ dispersion of 1000 bootstrap realisations. 
The profiles were estimated using a $B_3$ kernel of width $h=4 \hMpc$ in the radial coordinate, but similar results are obtained when using slightly different widths or shapes of the kernel.
}
\label{fig:radprof}
\end{figure}

A simpler view can be obtained by calculating the average radial density profile $\rho(r)$ around the $N_r$ centres. 
The resulting profile, shown in Fig.~\ref{fig:radprof}, has the same features as the 3D view: a high bump at short scales, and a clear peak at about the acoustic scale, with a maximum at $r_{\rm max} = 109.5 \pm 3.9\hMpc$. 
The error in $r_{\rm max}$ was estimated using bootstrap realizations \citep{lup93}. 
This scale gives the radius of the baryon acoustic shells, and it is therefore a good estimator of the acoustic scale in the sample. 
We also show in Fig.~\ref{fig:radprof} the radial profiles restricted to different regions of the sphere, to better characterize the anisotropy of the distribution. 
We define two cones with a width of $45^{\circ}$ with respect to the line-of sight in each direction (we call these `near' and `far' regions), and a `transverse' region covering the belt between the cones. 
We obtain for each of these regions qualitatively similar results. 
As expected, we see how the `near' and `far' subsamples are more strongly affected by observational effects, such as redshift-space distortions, which are more severe along the line of sight. 
In contrast, the result for the `transverse' subsample matches, within the errors, that for the full sphere.
It is interesting to note that the value of $r_{\rm max}$ is slightly larger for the  `far' sample than for the `near' one.

It is worth to emphasize that this approach would be impossible with any statistical BAO detection method used this far, since the spatial localization of the shells is completely lost in the correlation function or in the power spectrum, while the local nature of the wavelet approach has allowed us to identify the positions of the most representative structures in our sample. Moreover, we are
measuring the acoustic scale at positions selected for their low contamination from other structures, which is not the case when averaging over the full sample. In this way, we maximize the BAO signal, while minimizing the effect of signals coming from different large-scale structures.

\section{Discussion and conclusions}
\label{sec:conc}

In summary, we have designed a new method for the detection of baryon acoustic oscillations in the galaxy distribution 
and for the localization, in configuration space, of the structures responsible for them.
This method is based on the use of a specially designed wavelet applied directly on the density field. Our approach also relies on the use of two different tracers: one for the overall density distribution, and the other for the central overdensities of the baryon acoustic structures.

After testing the method with simulations, we applied this method to the detection of baryon acoustic structures in a sample drawn from the SDSS. In this case, we used galaxies from the `Main' catalogue to trace the overall density field, and galaxies from the LRG catalogue to trace the location of massive dark matter haloes. 
We clearly detect BAO in the sample providing a confirmation of the detection obtained previously using general two point statistics (the power spectrum and correlation function).  In fact, our approach provides an independent method for the detection.
Finally, we showed how this method allows us, through the use of $W_{\rm max}(\mathbf{x})$, to localize in configuration space the actual structures responsible for the BAO signal obtained.
This is a consequence of using a wavelet acting directly on the density field.
We illustrate the utility of this approach by showing the density distribution stacked around a set of centres known to show the BAO feature given their $W_{\rm max}$ value.

Recent works have proposed alternative methods to study the BAO based on wavelets \citep{xu10a, tia11a}.  In particular, \citeauthor{tia11a} use a Mexican hat wavelet function with two parameters, conceptually similar to ours. They use it to search for a peak in the two point correlation function of the `Main' SDSS sample, obtaining a detection with a $p$-value $p = 0.002$ (equivalent to $3.1\sigma$ in the Gaussian case). As in our case, this shows the utility of using the `Main' sample to reduce the shot noise in the calculation and to obtain significant detections. However, these works apply the wavelet to the measured two point correlation function, instead of directly to the density field. 
In this way, the use the capabilities of the wavelets to characterize accurately the BAO signal (in terms of radius and width), but they are not able to get any information about the localization of these structures in space.

The use of wavelets directly on the density field isolates valuable information about the baryon acoustic structures that is hidden in the standard two point statistics. In particular it gives us information, through the coefficients $W_{R,s}(\mathbf{x})$, to localize regions in the sampled volume giving the largest or lowest signal. We expect that this new method for studying BAO will be of much use for ongoing or planned surveys, such as the WiggleZ Survey \citep{dri10a}, the Baryon Oscillation Spectroscopic Survey \citep[BOSS,][]{eis11a}, or the Physics of the Accelerating Universe (PAU) Survey \citep{ben08a}, which will cover a much larger volume than studied here, and will explore higher redshifts.

\begin{acknowledgements}
This work has been supported by the European Research Council grant SparseAstro (ERC 228261), by the Spanish CONSOLIDER projects AYA2006-14056 and CSD2007-00060, including FEDER contributions, by the Generalitat Valenciana project of excellence PROMETEO/2009/064, and by the Estonian grants SF0060067s08 and ETF8005. P.A.M. was supported by the Spanish Ministerio de  Educaci\'on through a FPU grant, and by an ERC StG Grant (DEGAS-259586). 

We acknowledge the use of data from the MICE simulations, publicly available at http://www.ice.cat/mice.
We also acknowledge the use of public data from SDSS.
Funding for the SDSS and SDSS-II has been provided by the Alfred P. Sloan Foundation, the Participating Institutions, the National Science Foundation, the U.S. Department of Energy, the National Aeronautics and Space Administration, the Japanese Monbukagakusho, the Max Planck Society, and the Higher Education Funding Council for England. The SDSS Web Site is http://www.sdss.org/.
The SDSS is managed by the Astrophysical Research Consortium for the Participating Institutions. The Participating Institutions are the American Museum of Natural History, Astrophysical Institute Potsdam, University of Basel, University of Cambridge, Case Western Reserve University, University of Chicago, Drexel University, Fermilab, the Institute for Advanced Study, the Japan Participation Group, Johns Hopkins University, the Joint Institute for Nuclear Astrophysics, the Kavli Institute for Particle Astrophysics and Cosmology, the Korean Scientist Group, the Chinese Academy of Sciences (LAMOST), Los Alamos National Laboratory, the Max-Planck-Institute for Astronomy (MPIA), the Max-Planck-Institute for Astrophysics (MPA), New Mexico State University, Ohio State University, University of Pittsburgh, University of Portsmouth, Princeton University, the United States Naval Observatory, and the University of Washington. 
\end{acknowledgements}

\bibliography{ArnalteMur_baolet}

\bibliographystyle{aa}

\end{document}